\documentclass[referee]{raa}           
\usepackage{listings}
\usepackage{xcolor}
\usepackage{graphicx,times}
\usepackage{natbib}
\usepackage{amssymb,amsmath}
\bibpunct{(}{)}{;}{a}{}{,}

\usepackage[pagebackref=true]{hyperref}

\begin{document}

   \title{Mock Observations for the CSST Mission: CPI-C -- Targets for High Contrast Imaging}

 \volnopage{ {\bf 20XX} Vol.\ {\bf X} No. {\bf XX}, 000--000}
   \setcounter{page}{1}

   \author{Yi-Ming Zhu
   \inst{1,2}, Gang Zhao\inst{1,2}, Jiang-Pei Dou\inst{1,2}, Zhong-Hua Lv\inst{1,2}, Yi-Li Chen\inst{1,2}, Bo Ma\inst{3,4}, Zhao-Jun Yan\inst{5}, Jing Tang\inst{6}, Ran Li\inst{7,8}
   }

   \institute{Nanjing Institute of Astronomical Optics $\&$ Technology, Chinese Academy of Sciences, Nanjing 210042, China; {\it gzhao@niaot.ac.cn, jpdou@niaot.ac.cn}\\
        \and
             CAS Key Laboratory of Astronomical Optics $\&$ Technology, Nanjing Institute of Astronomical Optics \& Technology, Nanjing 210042, China\\
        \and
             School of Physics and Astronomy, Sun Yat-sen University, Zhuhai 519082, China\\
        \and
             Center of CSST in the great bay area, Sun Yat-sen University, Zhuhai 519082, China\\
        \and
             Shanghai Astronomical Observatory, Chinese Academy of Sciences, Shanghai 200030, China\\
        \and
             CAS Key Laboratory of FAST, National Astronomical Observatories, Chinese Academy of Sciences, Beijing 100101, China\\
        \and
             School of Physics and Astronomy, Beĳing Normal University, Beĳing 100875, China\\
        \and 
             School of Astronomy and Space Science, University of Chinese Academy of Sciences, Beĳing 100049, China\\
\vs \no
   {\small Received 20XX Month Day; accepted 20XX Month Day}
}

\abstract{We introduce \texttt{CPISM}, a simulation program developed for the Cool Planet Imaging Coronagraph (CPI-C) on the China Space Station Telescope (CSST). \texttt{CPISM} supports high-contrast exoplanet imaging by simulating observational conditions and instrumental effects to optimize target selection and observation strategies. The modular design includes target modeling, imaging simulation, observational effects, detector response, and data product generation modules, enabling flexible and realistic synthetic observations. Validation through simulations of a bright star shows strong agreement with theoretical expectations, confirming the program's accuracy. \texttt{CPISM}'s modular design allows flexibility, accommodating different stellar and planetary models, and can simulate instrumental noise, cosmic rays, and other observational effects. This tool aids in data processing, signal-to-noise ratio analysis, and high-contrast photometry, contributing to future exoplanet discovery and characterization efforts. The program’s outputs will enhance observation planning and scientific return for the CPI-C mission, providing critical insights into exoplanetary systems.
}

   \authorrunning{Y.-M. Zhu et al. }            
   \titlerunning{CPI-C target simulation}  
   \maketitle

%
\section{Introduction}           
\label{sect:intro}

Over the past few decades, the exploration of extrasolar planets has discovered more than 6,200 exoplanets orbiting different stars, of which dozens have been discovered by direct imaging methods (\citealt{2024CRPhy..24S.139C}).
High-contrast imaging has emerged as a crucial tool for directly observing exoplanets, providing invaluable data on planetary atmospheres, surface compositions, and orbital dynamics. Groundbreaking instruments like the Gemini Planet Imager (\citealt{2014PNAS..11112661M}) and Spectro-Polarimetric High-contrast Exoplanet REsearch (SPHERE, \citealt{2019A&A...631A.155B}) have already revolutionized our understanding of exoplanet systems by delivering high-resolution images that offer unprecedented contrast between stars and their orbiting planets (\citealt{2008Sci...322.1348M,2019A&A...632A..25M}). 

The \textit{Cool Planet Imaging Coronagraph} (CPI-C) aboard the \textit{China Space Station Telescope} (CSST) represents a major advancement in exoplanet research~\citep{2026RAA....26e5021D}. With its capability to achieve a contrast ratio of $10^{-8}$, CPI-C  is set to make significant contributions to our understanding of exoplanets in the Milky Way. CPI-C stands out due to its advanced coronagraph technology, which allows it to achieve exceptional contrast levels across both optical and infrared wavelengths. This capability is crucial for detecting faint exoplanetary signals near their host stars at separations ranging from $0.19$ to $1$ arcseconds. By combining these advanced features, CPI-C opens up new possibilities for studying the atmospheric composition of exoplanets and could play a key role in the search for biomarkers and signs of habitability.


Before CPI-C officially begins its observations, it is essential to make preliminary preparations through simulations. These preparations include the characterize and select of observation targets, as well as the development of the survey strategy.

Simulations play a critical role in optimizing future observations by enabling astronomers to refine target selection and observation strategies. Furthermore, by simulating instrumental effects and observational environments, simulations help mitigate risks and maximize scientific outcomes. Similar approaches have been employed for other exoplanet detection instruments, such as The Coronagraph Instrument (CGI) (\citealt{2022SPIE12180E..1XP}) on the Roman Space Telescope (RST) (\citealt{2015arXiv150303757S}) and The Pandora SmallSat (\citealt{2022SPIE12180E..0CH}), both of which have developed simulation programs (\citealt{2020SPIE11443E..38D,2024SPIE13101E..0FH}).

In this work, we introduce a program named \texttt{CPISM} (CPI-C Image Simulator) used to simulate the observation of CPI-C. \texttt{CPISM} connects multiple stages of the mission workflow — from target selection and observation configuration to data format generation and scientific evaluation — and serves as a unified interface between instrumental models, data processing pipelines, and science planning tools.

 The simulation platform supports both engineering validation and scientific exploration. It provides mock Level 0 data formatted for pipeline development, enables photometric analysis of simulated planets, and supports observation strategy optimization through contrast and signal-to-noise predictions. A companion paper (Zhao et al. 2025, in preparation) presents the detailed modeling of CPI-C’s optical system and detector physics. Here, we focus on the system-level design, architecture, and end-user applications of the \texttt{CPISM} platform.

The paper is structured as follows: Section \ref{sec:cpism} introduces the simulation program's module composition and operational steps. Section \ref{sec:sim_workflow} describes the process of simulating CPI-C observation targets using the program. In Section \ref{sect:example}, we verify the simulation program's performance for observing targets through an example. Finally, Section \ref{sec:dis} presents the summary and discussion.

\section{System Architecture} \label{sec:cpism}

\subsection{Design Overview}

To support both engineering and scientific preparation for CPI-C operations, we have developed a simulation \texttt{Python}-based software package named \texttt{CPISM}. This tool serves two major purposes:

1. To generate realistic Level 0 data that mimics raw CPI-C observations, including detector and instrumental artifacts, which can be used to develop and validate the mission’s data processing pipeline;

2. To provide synthetic observations for scientific research, including target detectability studies, signal-to-noise ratio (SNR) analysis, photometric precision tests, and observation strategy optimization.

The architecture is designed with flexibility and clarity in mind, featuring modular components, well-defined interfaces, and seamless integration with both scientific workflows and instrument-level simulations.

First, modularity is fundamental to \texttt{CPISM}’s structure. The simulation framework is composed of functionally independent modules—including target modeling, optical imaging, detector response, observational effects, and data product generation—each of which can be developed, tested, and replaced separately. This design facilitates focused development on individual components without compromising the integrity of the overall system.

Second, the platform is highly extensible. It is designed to incorporate external models with minimal modification. Users can replace the default planetary atmosphere or stellar spectrum generators with high-resolution alternatives, integrate new noise or stray light models, or adapt the simulation to accommodate updates in the CPI-C optical system and detector configuration.

Third, standardized data interfaces ensure seamless communication between internal modules and interoperability with external systems such as observation planning and data process pipelines. Simulation inputs—including target catalogs, instrumental settings, and observational parameters—are passed through structured dictionaries or configurable class instances. Output products include both a structured image array and FITS files conforming to CSST Level 0 data conventions.

Fourth, \texttt{CPISM} is developed for dual-purpose usability, serving both scientific and engineering needs. On one hand, it generates Level 0 FITS data products that mimic the raw output of CPI-C and enable realistic testing of the data reduction pipeline. On the other hand, the simulator supports science-driven applications such as signal-to-noise ratio (SNR) estimation, contrast curve evaluation, and target prioritization, thereby bridging technical verification with scientific optimization.

Finally, CPISM is designed for interoperability with detailed physical models. The architecture accommodates precomputed PSFs, EMCCD response functions, and other high-fidelity outputs from physical modeling as drop-in components. This capability allows \texttt{CPISM} to serve as a system-level wrapper that combines realistic hardware behavior with customizable science scenarios.

Overall, \texttt{CPISM} provides a flexible framework for generating synthetic observations with instrument- and detector-specific characteristics. It supports image-based SNR analysis, high-contrast photometry, astrometric measurement, and optimization of survey strategies for CPI-C. The \texttt{CPISM} simulation software is publicly available at the official CSST simulation website: \url{https://csst-tb.bao.ac.cn/simulation/cpic/installation.html}, where users can access installation instructions, usage documentation, and example datasets.

\subsection{Modular Structure}
The performance of the CPI-C depends on advanced optical components, including a deformable mirror, apodizing filter, and focal plane mask, which together achieve a contrast of $10^{-8}$. These components collectively suppress starlight and enhance the visibility of faint planetary companions. A key performance metric is the Point Spread Function (PSF), which defines the image quality. And  careful optimization ensures that the simulated PSF closely matches the performance expected from the satellite (Zhao et al. 2025, in preparation). CPI-C uses an Electron Multiplying Charge-Coupled Device (EMCCD) camera for optical imaging, which amplifies faint signals for improved sensitivity. These technologies collectively enable the detection and detailed study of exoplanets and their atmospheres.

The selection of CPI-C observation bands is guided by the characteristics of the planetary spectrum. Based on the absorption and platform characteristics derived from planetary spectrum simulations, four optical observation bands were designed, with central wavelengths of 520 nm, 662 nm, 720 nm, and 850 nm, as well as four infrared observation bands, with central wavelengths of 940nm, 1265nm, 1425nm, and 1542nm.

\begin{figure*} 
   \centering
   \includegraphics[width=15.0cm, angle=0]{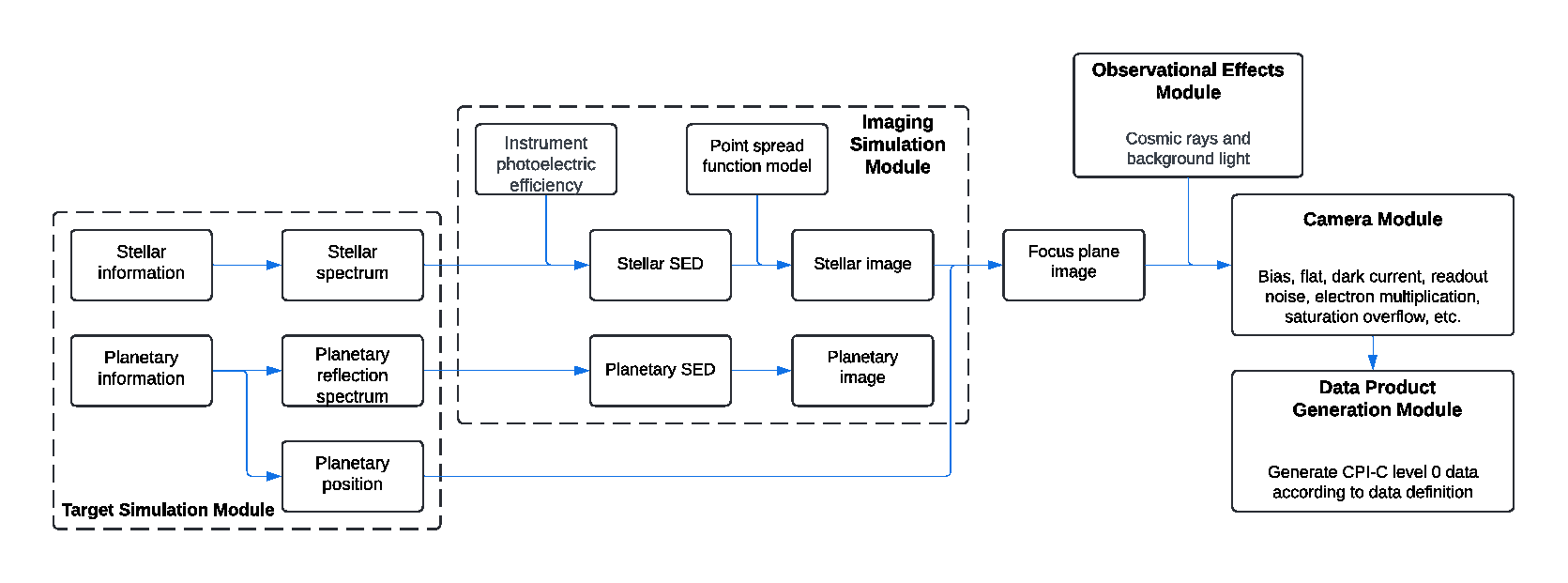}
   \caption{The working flowchart of the \texttt{CPISM} program.} 
   \label{fig:flowchart}
\end{figure*}

To support such a complex observational system, \texttt{CPISM} adopts a modular design that mirrors the structure of the actual CPI-C instrument and its operational pipeline. The simulation framework is organized into five major components: the Target Simulation Module, Imaging Simulation Module, Observational Effects Module, Camera Simulation Module, and Data Product Generation Module. Each module is responsible for simulating a distinct aspect, and the modules communicate via standardized internal data structures. As shown in Fig. \ref{fig:flowchart}, the framework simulates the full observational sequence of CPI-C, transforming astrophysical target parameters into realistic Level 0 data products. A brief overview of the module responsibilities is summarized as follows:

\begin{enumerate}
    \item \textbf{Target Simulation Module:} This module is responsible for generating the spectral and spatial characteristics of astrophysical sources. It models stellar spectral energy distributions (SEDs) based on the grid from \citet{castelli_kurucz_2004} and incorporates a planetary reflection spectrum model from \citet{2018AJ....156..158B} parameterized by metallicity and cloud properties. By incorporating the Lambertian phase model, contrast model, and stellar spectrum model, the program generates the reflected spectra of planets based on varying distances, radii, star-planet relative positions, and planetary physical properties.

    \item \textbf{Imaging Simulation Module:} This component models the instrument’s optical system, including the apodizing pupil filter, wavefront corrector, and focal plane mask. Using Fourier optics methods, it computes the system PSF and convolves it with the target scene to produce a simulated image at the focal plane. Dark zone regions with high-contrast suppression are constructed to emulate CPI-C’s expected performance. Although CPI-C operates in space and therefore does not need a conventional ground-based adaptive optics (AO) system to correct atmospheric turbulence, the coronagraph does contain an internal deformable mirror (DM) that runs in a closed-loop wavefront-control mode with an onboard wavefront sensor. This active control suppresses quasi-static aberrations and sustains a dark‑hole region required for high‑contrast imaging. In addition, the simulator includes a wavelength-dependent PSF generation module that models the imaging system’s spectral response across each CPI-C observation band.

    \item \textbf{Observational Effects Module:} To achieve realism, this module simulates background light and cosmic ray contamination. A cosmic ray model tailored to the CPI-C imaging camera was developed by referencing cosmic ray energy, frequency, and characteristics from tests conducted on the Hubble Space Telescope (HST) (\citealt{2021ApJ...918...86M,2024acsd.book...13H}). The methods are described in detail in the \ref{sub:obs_eff}. For background light, the solar spectrum was used to describe the background's spectral characteristics, resulting in a uniform background light model.

    \item \textbf{Camera Simulation Module:} This module simulates the response of CPI-C’s EMCCD detector, modeling the full readout chain from photon collection to final digital output. It incorporates key instrumental effects, including dark current, flat-field nonuniformity, nonlinearity, clock-induced charge (CIC), bad columns, vertical blooming, and electron multiplication. The EM gain is calculated as a function of voltage and temperature using calibration-based models. Bias structures such as striping, interference, and drift are also included. The output is a 2D image array consistent with the expected raw telemetry format.

    \item \textbf{Data Product Generation Module:} The final module assembles the outputs of the upstream modules into structured Level 0 data products. These products include FITS images formatted according to CSST conventions, with appropriate headers, metadata, and version control. The output hierarchy supports calibration, engineering, and scientific modes, and is compatible with downstream pipelines for data processing.

\end{enumerate}

By organizing the simulation pipeline into clearly defined modules, \texttt{CPISM} ensures maintainability, extensibility, and compatibility with both upstream physical models and downstream analysis tools. Detailed operational logic for each module is presented in Section \ref{sec:sim_workflow}.

\subsection{Level 0 Data Products}

\begin{figure}
   \centering
   \includegraphics[width=12.0cm, angle=0]{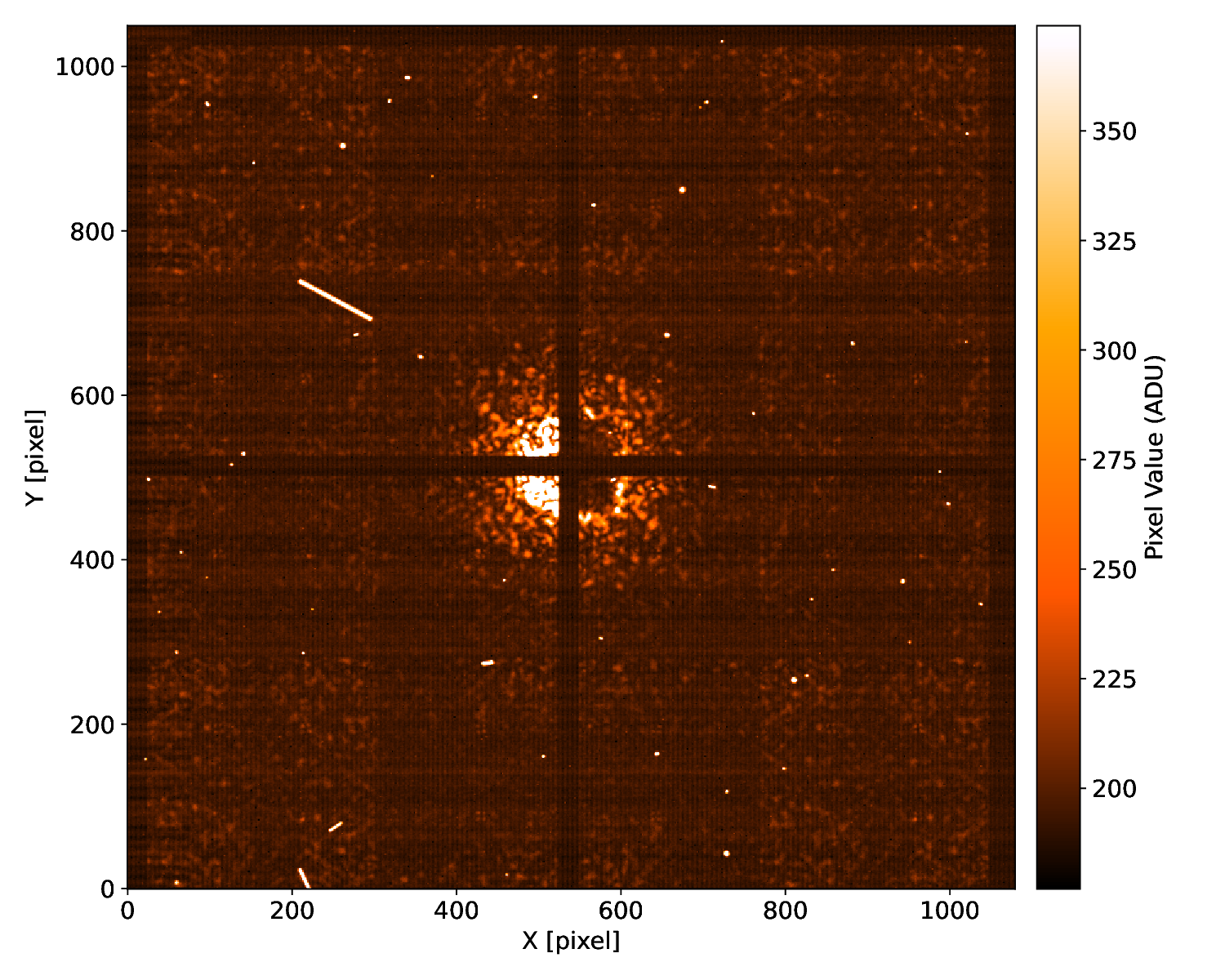}
   \caption{An image of CPI-C level 0 data simulated by the \texttt{CPISM} program. The frame contains 1080 × 1050 pixels. With the adopted pixel scale of 0.016153 $\mathrm{arcsec\, pix^{-1}}$, this corresponds to an on-sky footprint of 17.45 $\mathrm{arcsec}$ × 16.96 $\mathrm{arcsec}$. The color bar indicates the pixel value (analogue-to-digital units, ADU) on a linear scale.} 
   \label{fig:img}
\end{figure}

The simulation program generates CPI-C Level 0 data files, which consist of raw observational data captured by the optical and near-infrared cameras. These data are stored in the FITS (Flexible Image Transport System) format and include images, auxiliary data, calibration information, and engineering telemetry—all essential for subsequent data analysis and processing (see Fig. \ref{fig:img} for an example).
 
Each exposure in the CPI-C module produces one or more images containing pixel data that represents the light collected from observed targets, such as exoplanets and background stars, across various bands. These images are unprocessed at this stage and serve as raw input for further analysis.

The naming convention for the Level 0 data files follows a structured format to ensure clear identification of the observation and data type:

\begin{itemize}
    \item \textbf{File format}:

     \texttt{CSST\_CPIC\_CAMERA\_TYPE\_YYYYMMDDhhmmss\_YYYYMMDDhhmmss\_OBSID\_X\_L0\_VER.fits}
    \begin{itemize}
        \item \textbf{CAMERA}: Denotes whether the data was collected by the visible light (\texttt{VIS}) or near-infrared (\texttt{NIR}) camera.
        \item \textbf{TYPE}: Indicates the target type, with a total of eight types: five calibration modes — \texttt{BIAS} (background), \texttt{DARK} (dark field), \texttt{FLAT} (flat field), \texttt{BKG} (sky background calibration), \texttt{LASER} (internal laser calibration) — and three scientific target modes— \texttt{SCI} (scientific observation), \texttt{DSF} (dense star field observation), \texttt{CALS} (calibration star observation).
        \item \textbf{YYYYMMDD}: The observation date in year-month-day format.
        \item \textbf{hhmmss}: The UTC observation time in hour-minute-second format.
        \item \textbf{OBSID}: A unique observation identifier.
        \item \textbf{X}: A supplementary file name that includes no specific information; it's used for formatting purposes.
        \item \textbf{L0}: Indicates that the file contains Level 0 (raw) data.
        \item \textbf{VER}: Version number of the data file.
    \end{itemize}
\end{itemize}

\begin{figure}
   \centering
   \includegraphics[width=12.0cm, angle=0]{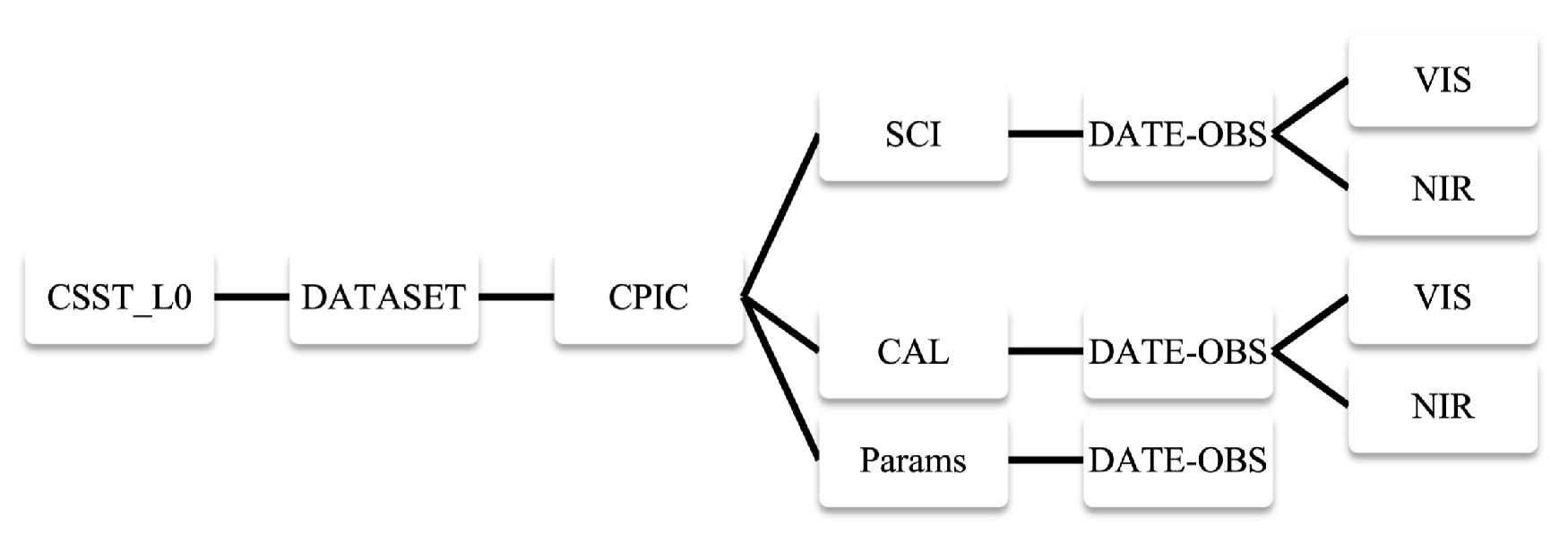}
   \caption{Schematic diagram of CPI-C Level 0 data storage folder structure.} 
   \label{fig:level0}
\end{figure}

Level 0 data files are organized in a hierarchical directory structure designed for easy access and categorization. The structure is divided into folders based on the type of data stored. The \texttt{SCI} folder contains the scientific observation data, including raw images from the telescope's cameras. The \texttt{CAL} folder holds calibration data, such as dark frames, flat field images, and background calibration, which are critical for removing instrumental effects and enhancing the quality of the scientific data. The \texttt{Params} folder contains engineering telemetry and instrumental parameters, providing the necessary context for accurately interpreting the observations. The folder naming hierarchy is shown in Fig. \ref{fig:level0}. This organizational structure facilitates the efficient retrieval of specific data types, whether for scientific analysis or for monitoring instrument performance.

When \texttt{CPISM} is executed for an observing sequence, it automatically constructs the hierarchical directory (e.g. \texttt{SCI/\textless MJD\textgreater} or \texttt{CAL/\textless MJD\textgreater}). All frames ($N$ images) acquired in that sequence are written to a single multi-extension FITS Level 0 file: a primary HDU followed by one \texttt{IMAGE} extension per frame (\texttt{HDU\,1}\dots\texttt{HDU\,$N$}). Separate observing sequences, each identified by a distinct \texttt{OBSID}, are saved as additional files and are placed automatically in the same directory tree.

By generating these Level 0 data files, the simulation program supplies the essential raw material for subsequent stages of data processing, ensuring that high-quality scientific data can be derived from the simulated CPI-C observations.

\subsection{Extensibility and Integration}

\texttt{CPISM} is architected with extensibility as a core design principle, enabling users to adapt, upgrade, or replace key physical models as the instrument evolves or new scientific requirements emerge. Its modular Python implementation, coupled with structured configuration interfaces, allows for seamless integration of external components and high-fidelity simulations.

In the target simulation module, stellar spectra are generated using the model grid from \citet{castelli_kurucz_2004} (hereafter CK04) by default. However, the spectrum interpolation framework—implemented via indexed lookup and trilinear interpolation—is decoupled from the model source, allowing for replacement with alternative spectral libraries such as PHOENIX~\citep{2013A&A...553A...6H} or empirical SEDs. Similarly, the planetary reflection model, currently based on a \citet{2018AJ....156..158B} parameterization, can be replaced with more sophisticated atmosphere simulators such as PICASO, enabling the simulation of high-resolution albedo spectra and phase-dependent features.

The cosmic ray model within the camera module is initialized from empirical maps derived from Hubble Space Telescope data~\citep{2021ApJ...918...86M,2024acsd.book...13H}. The design supports substituting this model with alternative particle energy spectra and morphological templates, including those adapted to CPI-C’s specific orbit and shielding conditions.

On the instrument side, the camera module supports external calibration inputs for dark current, CIC maps, flat fields, and bad columns. Users may supply their own calibration products via configuration. The PSF kernels, EM gain response, dark current evolution, and optical aberration maps can be directly incorporated into CPISM as external data products. 

This tight coupling between system-level simulation and physical modeling ensures that \texttt{CPISM} remains a consistent and up-to-date proxy for CPI-C’s performance throughout its development and operations. Beyond its engineering role, \texttt{CPISM} can also serve as a scientific simulation engine for generating observation catalogs and testing observation strategies in batch mode.

\section{Simulation Workflow}
\label{sec:sim_workflow}

The simulation workflow integrates several key modules to produce the final output (see Fig. \ref{fig:flowchart}). The process begins by inputting the necessary parameters for stars and planets, including their physical properties and observational settings. Using models for stellar spectra, planetary reflection spectra, and contrast, the target simulation module calculates the corresponding spectral outputs.

Next, the imaging simulation module processes these inputs to simulate the focal plane image. It accounts for system characteristics such as the PSF, optical aberrations, and the modulation of the deformable mirror. By convolving the spectral information of the star and planet with these system parameters, the module generates a realistic focal plane image, incorporating the dark hole contrast (Zhao et al. 2025).

Subsequent modules introduce observational effects, including background light and cosmic ray artifacts, and simulate the response of the CPI-C EMCCD detector.  The result is a comprehensive simulation that produces Level 0 data products, fully integrating observational effects into the final image.

\subsection{Input Configuration}

\texttt{CPISM} adopts a structured and configurable input interface to accommodate a wide range of observational scenarios and model customizations. Simulation parameters are provided through Python dictionaries or YAML-based configuration files, which define the properties of stellar and planetary targets, observational settings, and instrument parameters.

The instrument configuration file defines camera-level parameters such as exposure time, electron multiplication settings, readout noise, full-well capacities, and a comprehensive set of effect toggles grouped under the switch dictionary. Each camera effect can be individually enabled or disabled. These flags allow users to simulate either idealized or fully realistic detector responses. Optional reference files for flat-field maps, dark frames, CIC maps, and bad columns are automatically loaded based on the default paths or user definitions.

The observation parameter file defines the physical and observational properties of the simulated scene. It specifies the central star's coordinates, spectral type, magnitude, and distance, as well as parameters for planetary and background objects, including radius, albedo model, phase angle, angular separation, and spectral model. Each object can utilize either a blackbody or an atmospheric spectrum. Default stellar and planetary reflection spectra are provided, but users may override these with custom data or analytical models to test alternative physical assumptions. Additionally, observational settings such as bandpass, frame number, sky background brightness, and spacecraft attitude can be configured.

\subsection{Target and Spectrum Modeling}
The target’s position in the program can be given in two ways: (1) by using the target’s equatorial coordinates, or (2) by specifying the position relative to the host star. 

When using equatorial coordinates to assign the target's position, the equatorial coordinates of the host star must also be provided. When specifying the target’s position relative to the primary star, both the angular separation (in arcseconds) and the position angle (in degrees) must be provided. In the program, the position angle is defined as increasing east of celestial north: $0^{\circ}$ corresponds to celestial north and increases counter-clockwise on the detector.

The program uses the \texttt{spectrum\_generator} function to read the parameters of the targets and generate the positions and spectra of all targets.

The input parameters are provided as a dictionary. The keys include \texttt{name} for specifying the target name, \texttt{cstar} for specifying the central star, \texttt{stars} for the background star list, and \texttt{planet} for the planet list.

For the host star, the required inputs are magnitude, spectral type, right ascension (RA), and declination (DEC). For background stars, the magnitude, spectral type, and either RA and DEC or their separation and position angle relative to the host star must be provided. For the planet, the required inputs include radius, albedo model parameters, phase angle, and either RA and DEC or separation and position angle relative to the host star.

After inputting these parameters, the program will generate the target’s position and spectrum based on the provided data. The output is a table containing the positions and spectra of each target.

The function \texttt{star\_photlam} in the program is used to generate stellar spectra. When utilizing this function, the target’s apparent magnitude, spectral type, and choice between a blackbody spectrum or a more realistic model spectrum (CK04) must be specified. 

The spectral interpolation function of this code is designed to generate high-fidelity stellar spectra based on three fundamental stellar parameters: effective temperature (\texttt{Teff}), metallicity (\texttt{metallicity}), and surface gravity (\texttt{log\_g}). The process begins with reading a pre-compiled CK04 catalog stored in FITS format, which contains indexed stellar spectra characterized by different parameter combinations. To optimize computational efficiency, the catalog is cached to prevent redundant I/O operations. Given a specific set of input parameters, the code identifies the eight closest spectra in the parameter space by iteratively filtering the catalog based on \texttt{Teff}, \texttt{metallicity}, and \texttt{log\_g}.

\begin{figure*} 
   \centering
   \includegraphics[width=15.0cm, angle=0]{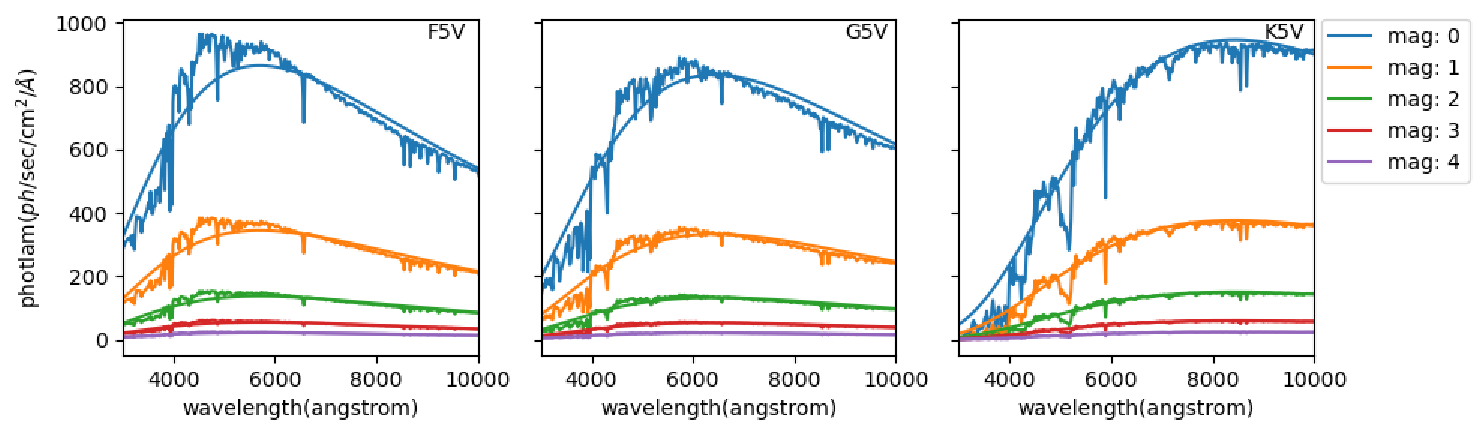}
   \caption{The stellar spectra of different spectral types (F5V, G5V, K5V) and their corresponding blackbody spectra across varying V-band magnitudes are generated by the program. } 
   \label{fig:stellar}
\end{figure*}

Once the relevant spectra are selected, trilinear interpolation is performed in three sequential steps. The final spectrum is derived through successive interpolation along \texttt{log\_g}, \texttt{metallicity}, and \texttt{Teff}, ensuring smooth transitions across parameter space. The resulting wavelength and flux data are stored in the \texttt{pysynphot} spectrum class, with wavelengths in angstroms (\AA), and the flux in \textit{photlam} (ph/sec/cm$^2$/\AA). Fig. \ref{fig:stellar} presents example spectral plots for various star types and V-band magnitudes.

\begin{figure} 
   \centering
   \includegraphics[width=12.0cm, angle=0]{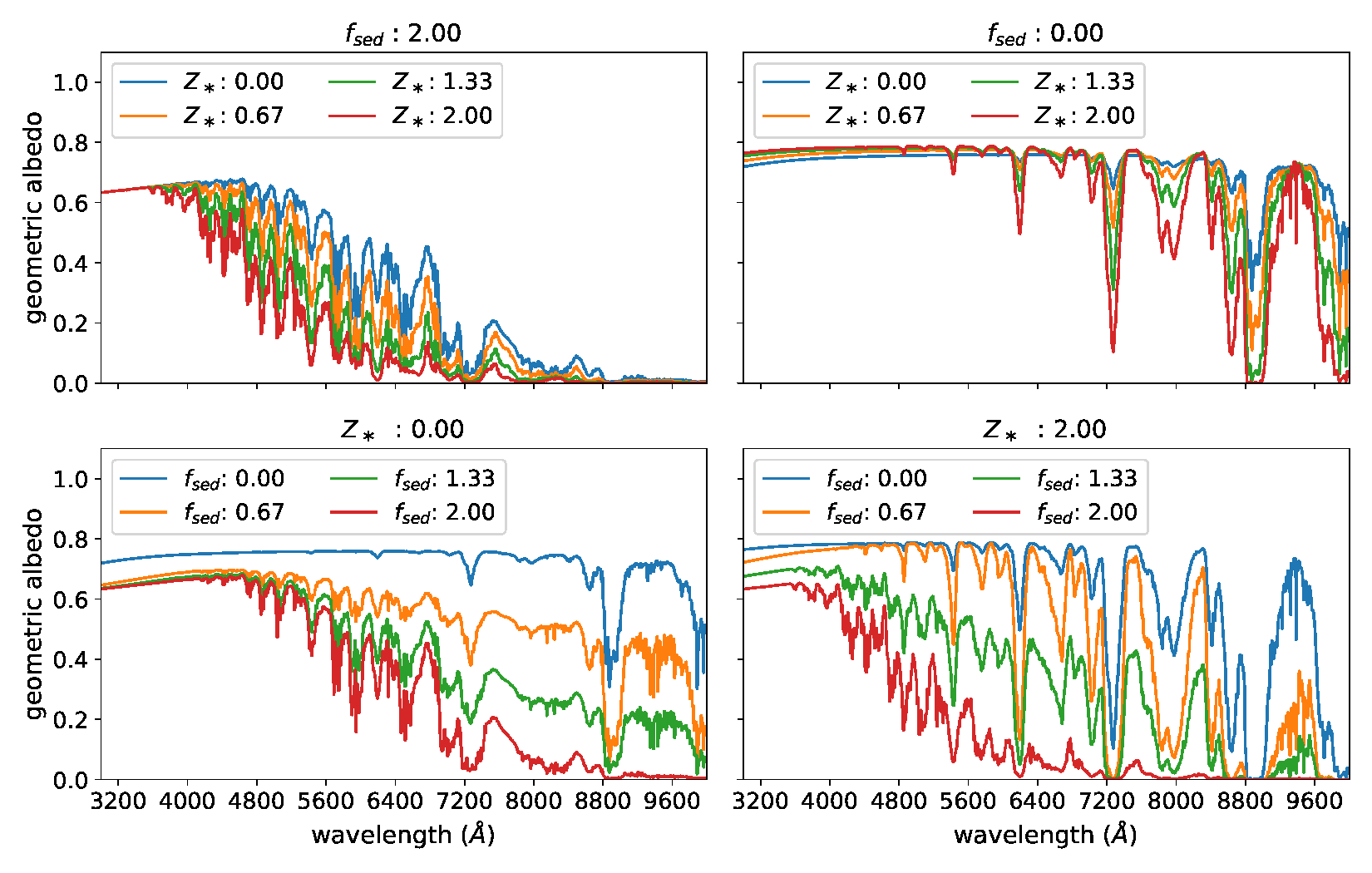}
   \caption{The the geometric albedo spectra of planets with difference atmosphere parameters. Top: Spectra of different metallicity parameters when the sedimentation efficiency parameter is constant. Right: Spectra of different sedimentation efficiency parameters when the metallicity parameter is constant.} 
   \label{fig:planetsed}
\end{figure}

The simulation program includes a built-in planetary reflected model from \citet{2018AJ....156..158B}. The primary factors influencing planetary spectra include methane and cloud formations. Two key parameters affecting the brightness profile are examined: the abundance of absorbing gases, represented by metallicity ($Z_*$) within the range of (0, 2), and cloud height and scattering properties, represented by sedimentation efficiency ($f_{\rm sed}$) in the range of (-2,2). A higher $f_{\rm sed}$ indicates more efficient sedimentation, leading to thinner clouds.
Fig. \ref{fig:planetsed} demonstrats the functionality of this model.

\subsection{Imaging and Contrast Simulation}

The optical path of CPI-C is modeled using Fourier optics. The module first constructs a 2D electric field across the pupil plane, optionally modified by an apodizing filter. This filter is represented as a grayscale transmission map , modulating the amplitude of the incident wavefront across the telescope aperture. The apodization function is handled using \texttt{HCIPy}-compatible arrays.

Next, a wavefront correction surface—typically derived from a deformable mirror (DM)—is optionally applied to the complex field to simulate quasi-static aberration control. Although \texttt{CPISM} does not implement closed-loop wavefront optimization internally, precomputed DM shapes or PSF kernels (from Zhao et al. 2025) can be loaded via the function \texttt{load\_psf()} and injected into the pipeline.

The modified electric field is then propagated to the focal plane using FFT-based angular spectrum methods. A focal plane mask may be applied to suppress diffraction structures near the star. This combination of optical components produces a structured PSF, including the dark hole—a region of high suppression, typically from $4-15 \lambda/D$.

Once the wavelength-dependent PSF has been constructed, the simulator proceeds to synthesize the focal plane image by combining the modeled astrophysical scene with the optical response. Each astrophysical source—whether a central star, planet, or background object—is represented by its spectral flux within the selected observation band and its angular position relative to the field center. These sources are projected into a spatial flux map according to their brightness and sky location.

To generate the final image, the simulator performs a convolution between the flux map and the PSF kernel associated with the current bandpass. In practice, due to computational efficiency considerations and because planetary and background sources are faint (making subtle off-axis PSF variations negligible), the simulator adopts a simplified convolution-based strategy. Two types of PSFs are generated for each observing band: an on-axis PSF that includes the cross-shaped focal-plane mask and represents the central star, and an off-axis, unmasked PSF representing all other targets. For off-axis sources, an idealized (delta-function) flux map is first created, placing point sources at fractional-pixel positions on a finely-sampled grid (typically 4×4) to ensure sub-pixel positional accuracy. Extended sources are similarly placed via interpolation. This idealized flux map is then convolved with the off-axis, unmasked PSF. Finally, the central star—represented by the masked PSF—is added. This hybrid treatment balances realism and computational efficiency, and has minimal impact on accuracy, as the faintness of off-axis sources renders minor PSF variations imperceptible.

This process captures the effects of diffraction, wavefront modulation, and focal plane suppression, resulting in a realistic intensity distribution across the field. The convolution step uses either precomputed kernels or numerically generated PSFs, and it accounts for key imaging parameters such as pixel scale, oversampling factor, and total field-of-view. The output is a simulated focal plane image that includes the main stellar PSF, planetary companions with reduced brightness, and other background sources, all modified by the high-contrast optics of CPI-C.

In parallel with image synthesis, \texttt{CPISM} calculates the theoretical contrast of the planet relative to the host star  using the function \texttt{planet\_contrast}. This function requires the input of the planet’s radius (in units of Jupiter radii), the planet's separation from the host star in the RA and DEC directions (in AU), and the planet’s phase angle. The phase angle refers to the angular separation between the line of sight to the planet and the line connecting the planet to the host star. When the line of sight is aligned with the star-planet line, the phase angle is 0 degrees. When the line of sight is perpendicular to the star-planet line, the phase angle is 90 degrees. The program calculates the 3D physical separation between the planet and the star based on the input projected distance, using the Lambert function.

The planetary contrast is then calculated using the equation (\citealt{2012ApJ...747...25M}):

\begin{equation}
\text{contrast}_p = \frac{f_{\text{planet}}}{f_{\text{star}}}=A_g \left( \frac{R_p}{r} \right)^2 \left[\frac{\sin (\theta)+(\pi-\theta) \cos (\theta)}{\pi}\right],  \label{eq:con}
\end{equation}
where $A_g$ is the geometric albedo of the planet, $\theta$ is the phase angle, $R_p$ is the radius of the planet, and $r$ denotes the distance between the planet and the star. Note that the phase angle cannot be 0 degrees or 180 degrees.

\subsection{Observational and Instrumental Effects} \label{sub:obs_eff}

Cosmic rays were observed to have a wide range of morphologies, from point-like to elongated shapes (\citealt{2015JInst..10C8006F}), with an average energy deposition of approximately 2700 electrons per cosmic ray. \citet{2021ApJ...918...86M} also provided insights into the path lengths of cosmic rays, which were typically proportional to their angle of incidence, with an estimated average path length of about 200 micrometers.

For the purpose of evaluating the robustness of our image processing pipeline against high-energy particle interference, we implemented a synthetic cosmic ray (CR) frame generator based on physical and statistical models in the simulation. Cosmic rays interacting with camera sensors produce linear, streak-like artifacts that can corrupt scientific measurements, especially in space-based observations.

In order to simulate the real cosmic ray situation as much as possible, the energy distribution of the cosmic rays in the program adopts the detected results of The HST (\citealt{wfpc2handbook}). The incident particle rate is based on the results of \citet{2021ApJ...918...86M}, the peak of which is about $1 \,\rm particle/s/cm^2$. This rate was modulated by solar activity, with variations corresponding to the 11-year solar cycle.

For a given exposure time $t$ and image size (in pixels), the expected number of CR events is estimated as:

\begin{equation}
N = t \cdot A \cdot R_{\mathrm{cr}}, \label{eq:cr}
\end{equation}

where $A$ is the effective sensor area in $\mathrm{cm}^2$ and $R_{\mathrm{cr}}$ is the CR rate.

Each cosmic ray streak is characterized by a random length, width, deposited charge, and orientation angle. The length is derived from a simplified 3D track projection model, considering the detector depth and pixel pitch. The width is modeled as a Gaussian spread with random variation, and the deposited charge follows an exponential distribution modulated by a power-law tail.

To construct the two-dimensional cosmic ray artifacts, two models are employed. In the default mode, the CR track image is synthesized using a combination of Gaussian profiles and modulated Fourier-based noise to emulate natural variations in trail structure. Alternatively, a Monte Carlo-based model is adopted, where cosmic ray events are generated using a pre-defined set of track templates drawn from real or simulated data. These templates capture the empirical distribution of track sizes and morphologies, enabling more physically realistic CR simulations.

The resulting streak images are randomly positioned on the sensor frame, rotated according to the sampled angle, scaled by the corresponding flux, and overlaid with attention to edge blending and non-negativity. This method ensures both statistical consistency and structural diversity of the simulated CR artifacts.

\begin{figure*} 
   \centering
   \includegraphics[width=0.45\textwidth]{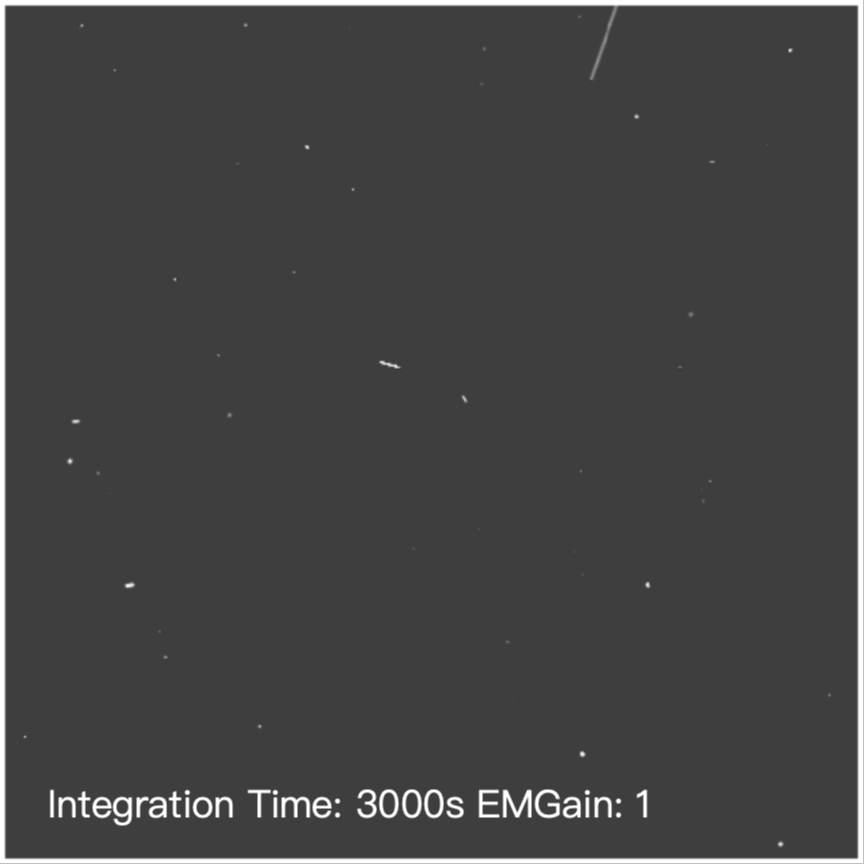}
   \includegraphics[width=0.45\textwidth]{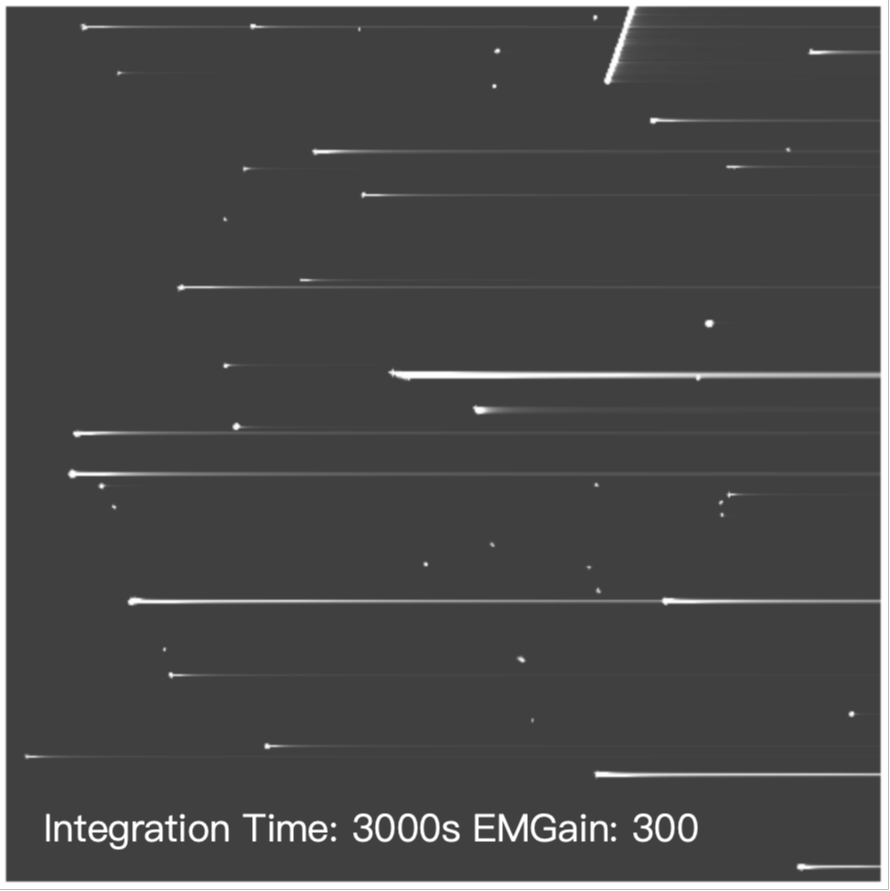}
   \caption{Simulated 3000-second exposure with EM gain of 1 (left) and 300 (right) for cosmic ray observation effect. When EMCCD uses high electron multiplication, cosmic rays may cause saturated overflow, resulting in a ``tailing'' phenomenon in the image.} 
   \label{fig:cr}
\end{figure*}

The effect of cosmic rays in the simulated observation image is shown in Fig. \ref{fig:cr}.

We also take background light into account in the simulation. Since the actual situation may be very complicated and the influence of background light is weak, we usually only add light with a flux density of $21 \,\rm mag/arcsecond^2$ uniformly to the image. And we reserve interfaces for zodiacal light and earth-atmosphere radiation, which can be modified in the future according to actual specific conditions. For stray light, users can also pre-calculate the corresponding stray light intensity data based on the specific characteristics of the optical system, and then set it in the program as an input parameter.

Detector effects intrinsic to the EMCCD are modeled comprehensively within the camera module. The simulation covers the entire EMCCD readout process, from photon conversion to final digital output. The photon collection phase converts photon flux into electrons, considering non-uniform pixel response and adding dark current according to predefined reference frames. Nonlinear response to illumination is modeled uniformly across the sensor array. When collected electrons exceed the pixel full-well capacity, vertical saturation and blooming occur, simulating the overflow behavior typical of EMCCDs.

The electron multiplication stage, characteristic of EMCCD technology, significantly amplifies incoming signals. Electron multiplication gain (EM gain) is implemented through a calibrated voltage-dependent relationship that also accounts for detector temperature. Effects specific to EMCCD operation, including clock-induced charge (CIC), readout noise, and charge transfer inefficiency, are explicitly simulated. Additional image artifacts, such as vertical striping, random horizontal interference patterns, and bias level drift, are included based on empirical detector characterization. These effects are parameterized in the camera configuration, allowing the simulator to flexibly represent idealized or realistic instrument behaviors. The culmination of these steps produces raw observational data formatted as Level 0 FITS files, closely aligned with anticipated real detector outputs.

\section{Example verification}\label{sect:example}

In the following section, we demonstrate the use of the  \texttt{CPISM} simulation program through an example of simulating a CPI-C target and verifying its accuracy.

\subsection{PSF chromaticity}

To evaluate the imaging performance of CPI-C across its designated observation bands, we simulate the system’s PSF and corresponding contrast curves at four representative wavelengths: 520 nm, 662 nm, 720 nm, and 850 nm. These simulations are generated using the \texttt{single\_band\_psf()} function in the \texttt{CPISM} framework, which constructs the optical field through Fourier optics and includes the effects of the apodizing pupil mask, wavefront correction, and focal plane mask. The dark zone is defined within a specific angular range where starlight suppression is optimized.

\begin{figure*} 
   \centering
   \includegraphics[width=\textwidth]{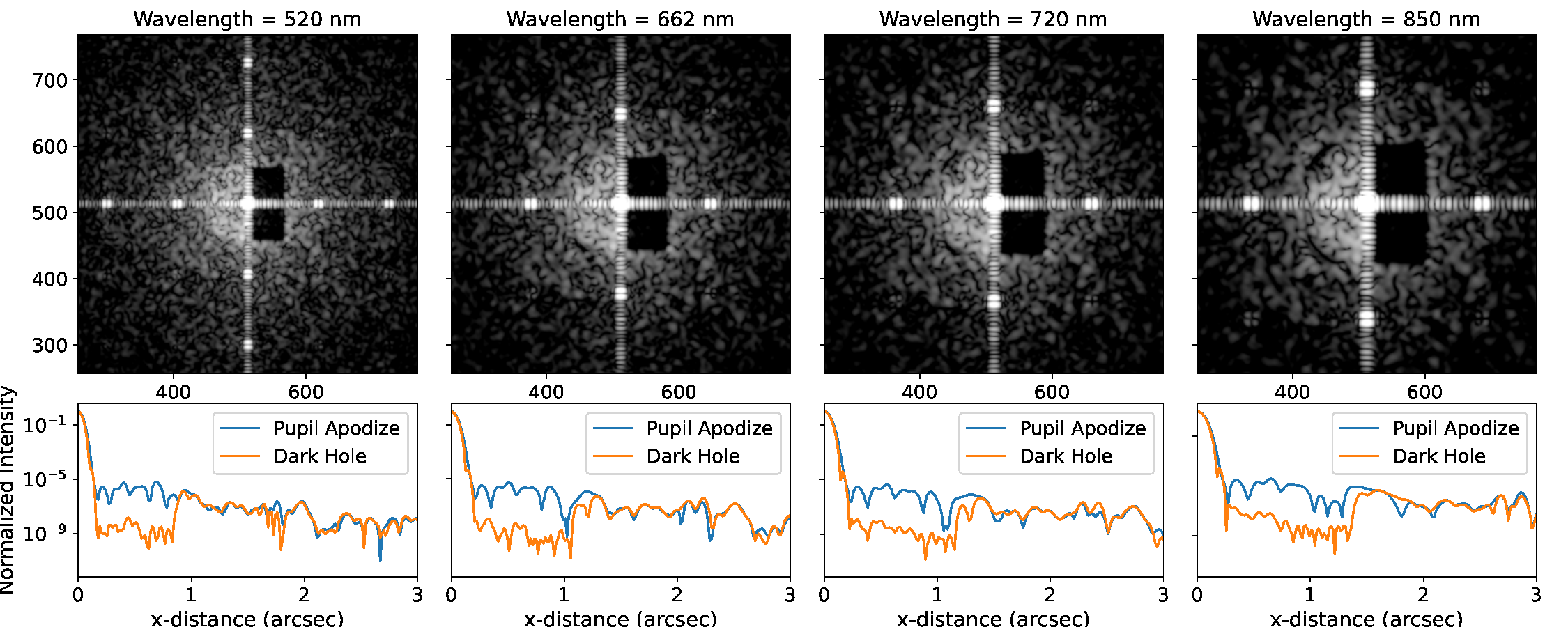}
   \caption{The PSF maps (top row) and the normalized contrast intensity profiles along the diagonal direction (bottom row) for each wavelengths. The orange and blue curves represent the intensity passing through or not passing through the dark hole, respectively.}
   \label{fig:psfcurve}
\end{figure*}

Fig.\ref{fig:psfcurve} presents the PSF maps (top row) and the corresponding normalized intensity profiles. Two imaging configurations are compared: one employing only pupil apodization, and the other incorporating the dark hole structure. The contrast curves are plotted on a logarithmic scale as a function of angular separation from the central star.

The simulation results show that the coronagraph consistently suppresses stellar light within the dark hole region (0.19 to 1 arcseconds) to below $10^{-8}$ in all four bands, with slight wavelength-dependent variations. The suppression is most effective near 662 nm, consistent with the optimized design of CPI-C. The dark hole region appears as a well-defined area of reduced intensity, clearly visible in both the PSF images and contrast curves. These results demonstrate that \texttt{CPISM} can reproduce the expected high-contrast imaging performance of CPI-C across multiple bands and provide a useful tool for validating the optical design and informing scientific analysis.

\subsection{Simulation of a Hypothetical Exoplanet}

CPI-C targets FGK-type stars within 40 pc of the Sun’s neighborhood for high-contrast imaging observations of potential exoplanets. The program uses multi-band photometry and spectral fitting to derive the atmospheric properties of detected exoplanets, laying the groundwork for future biosignature identification in exoplanetary systems. Based on the engineering specifications of CPI-C, there are nearly seven hundreds nearby stars selected as observation targets, most of which have V-band magnitudes between 0 and 7. Information about these stars, including spectral type, V-band magnitude, and ecliptic coordinates (for observation planning), was obtained from the SIMBAD database (\url{https://simbad.cds.unistra.fr/simbad/}). 

As an example, we selected Alpha Centauri (HIP 71681), a K1V-type star with a V-band magnitude of 1.35 (\citealt{1997ESASP1200.....E}), located at a distance of 1.35 pc (\citealt{2007A&A...474..653V}). A hypothetical planet is generated with the following parameters: position angle of $315^\circ$, phase angle of $90^\circ$, radius of $1\,R_{\rm Jupyter}$, and an angular distance of $0.65^{\prime \prime}$ from the host star. The metallicity is set to $Z_* = 1$ and the sedimentation efficiency to $f_{\rm sed} = 1$. Additionally, a background star is added for flux calibration, with a position angle of $225^\circ$, a spectral type of K1V (the same as the central star), and an angular separation of $0.95^{\prime \prime}$.

We use the simulation program to generate two sets of images: one containing the central star, planet, and reference star, and the other containing only the central star. Other parameters, such as exposure time, remain the same for both sets.

\begin{figure*} 
   \centering
   \includegraphics[width=0.3\textwidth]{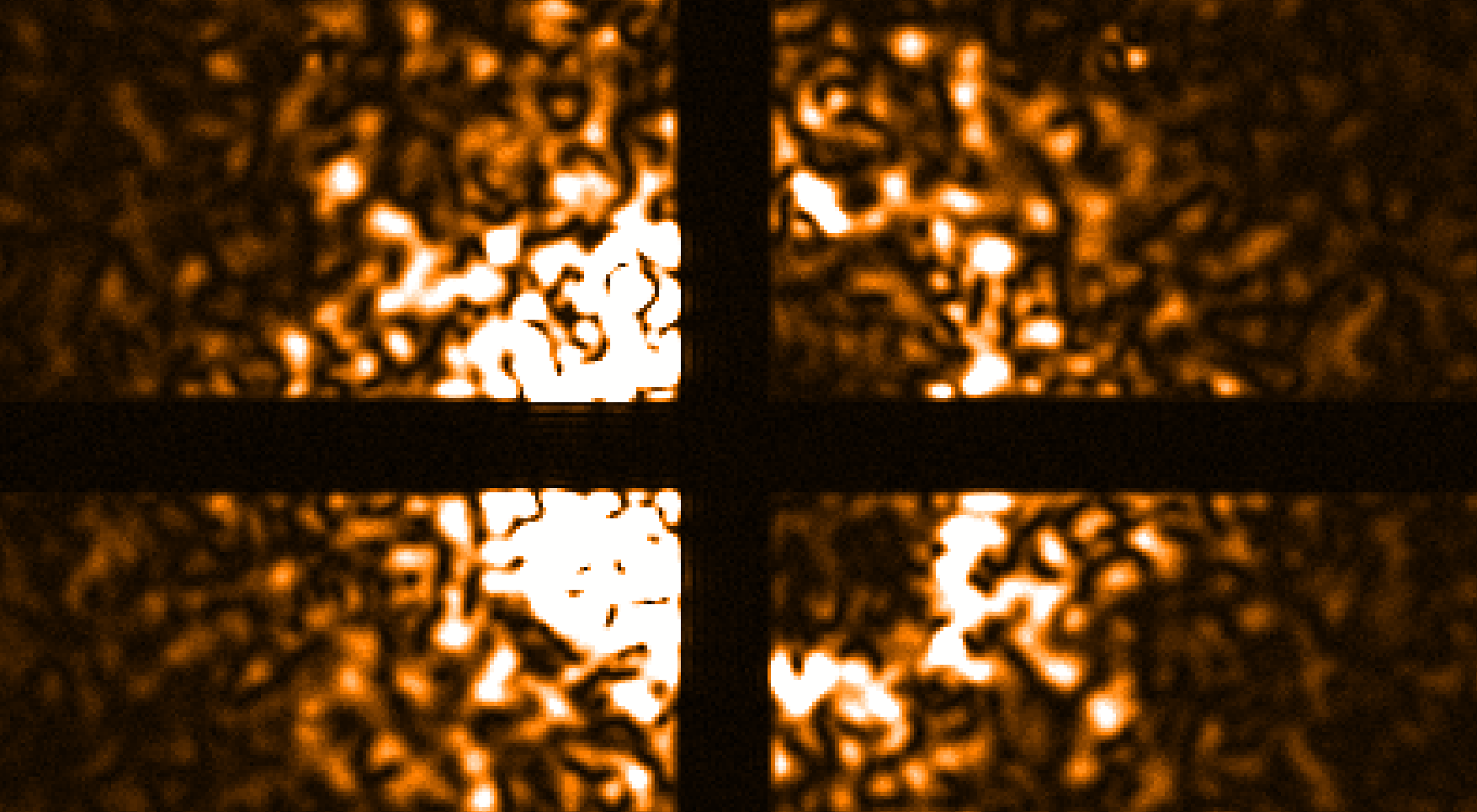}
   \includegraphics[width=0.3\textwidth]{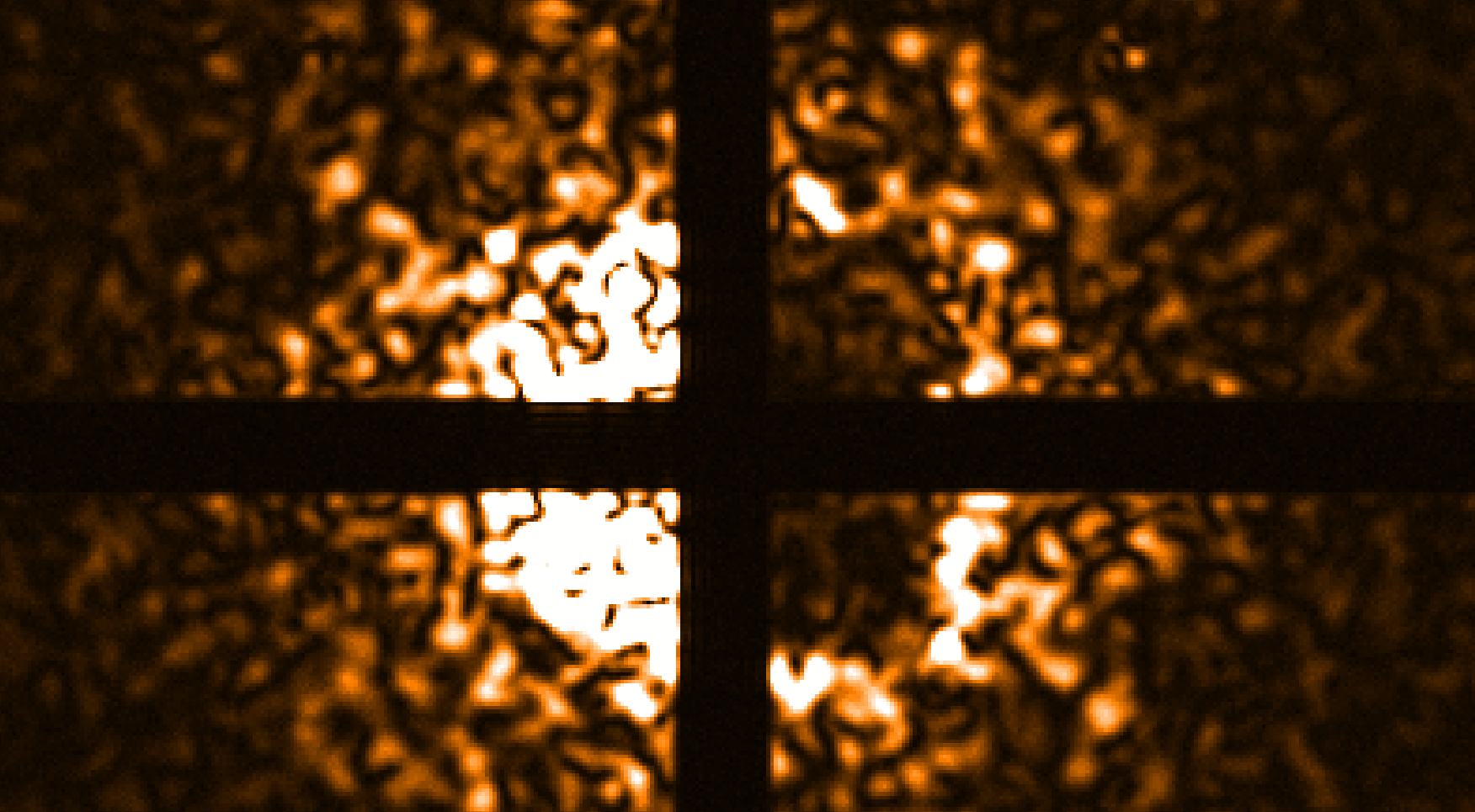}
   \includegraphics[width=0.3\textwidth]{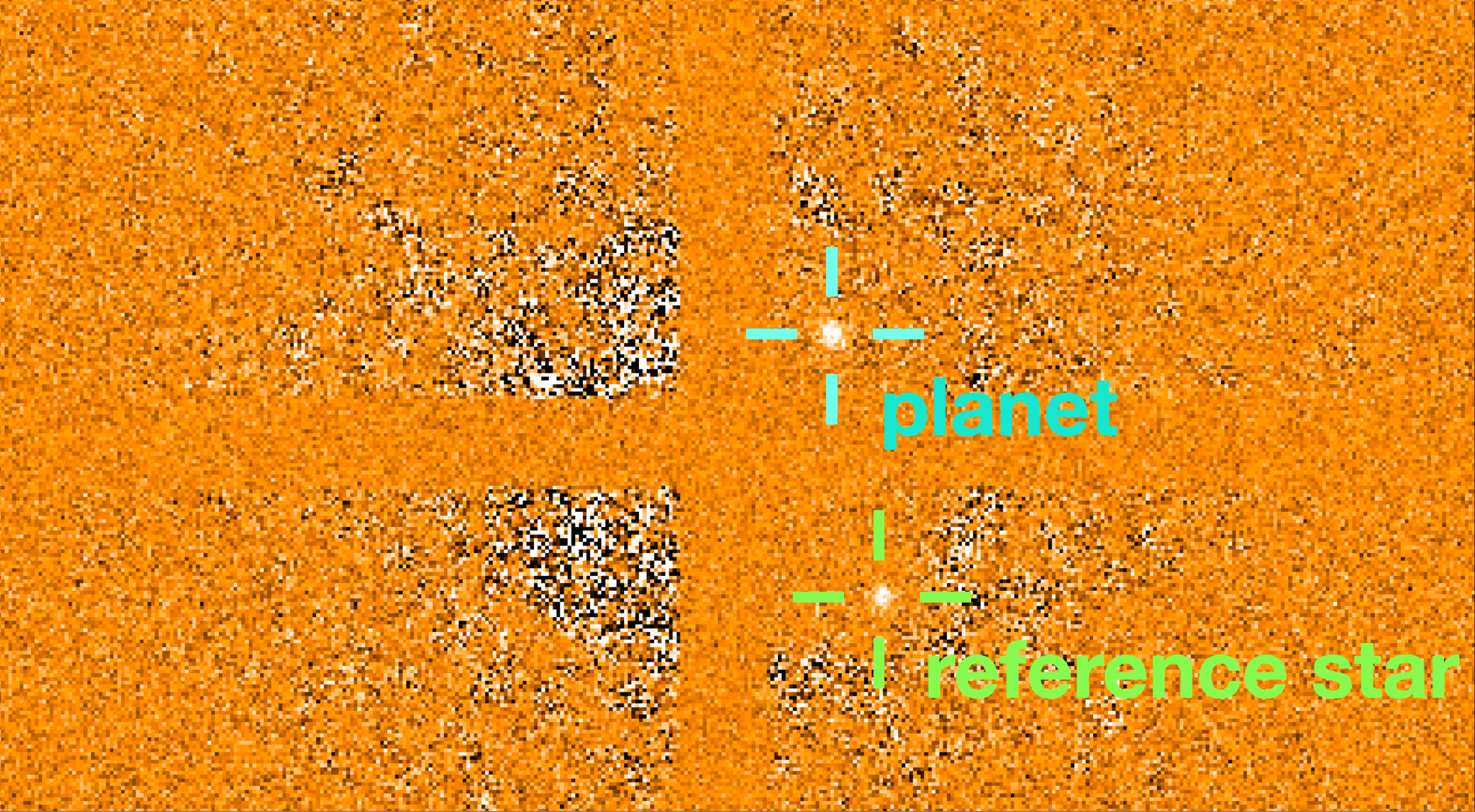}
   \caption{Left: The image contains the central star, planet, and reference star. Middle: The image contains only the central star. Right: The image after subtraction, where the planet and the reference star can be clearly seen. The simulated band is F520. The two crosses mark the positions of the planet and the reference star, which are located in different dark holes with position angles of $45^\circ$ and $135^\circ$ respectively.} 
   \label{fig:520sub}
\end{figure*}

After running the simulation, we adopt a Reference Differential Imaging (RDI) technique to suppress background noise in images containing planetary signals. For each science frame, its dark hole is extracted and linearly fitted using a library of background images known to be free of planetary signals. Singular Value Decomposition (SVD) is employed to construct an optimal background model specific to each frame~\citep{2021MNRAS.502.2158R}. Then the images containing only the central star are then subtracted from the images containing the planet and reference star. The subtracted frames are subsequently co-added to further reduce random noise, resulting in an image that includes only the planet and reference star, while largely removing the background. Fig. \ref{fig:520sub} shows an example of simulated F520-band images before and after subtraction. This simulation and image processing are conducted across all four observation bands of CPI-C.

\begin{figure*} 
   \centering
   \includegraphics[width=\textwidth]{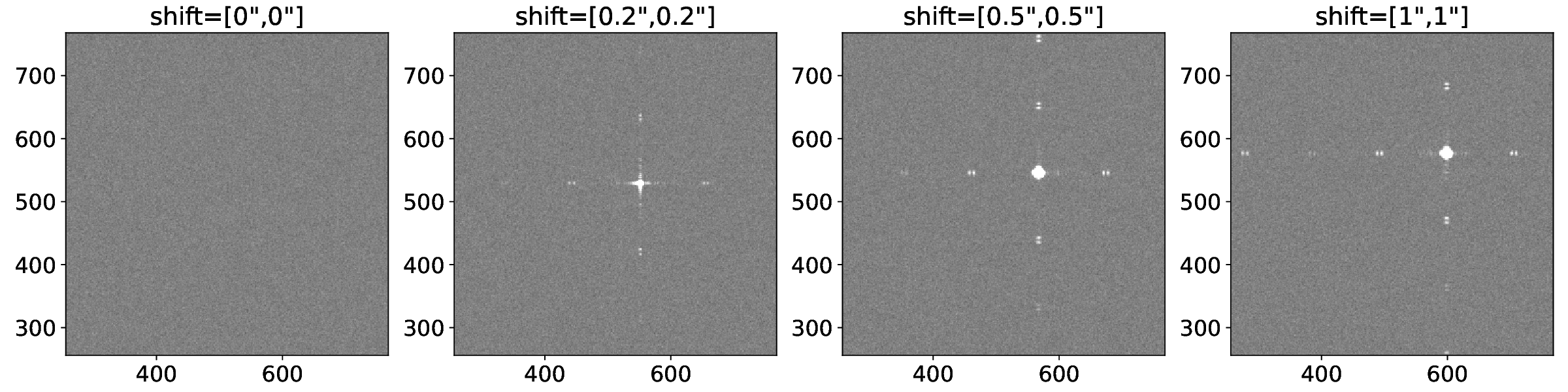}
   \caption{Simulated images of the star shifted away from the central obstruction. The images from left to right show the stars with no shift, a shift of $0.2^{\prime \prime}$, $0.5^{\prime \prime}$, and $1^{\prime \prime}$.Because central stars are typically very bright, multiple short-exposure images are taken and subsequently combined to avoid saturation. The simulation is conducted in the F520 band with the exposure time of $0.01\, \mathrm{s} \times 50$ for each situation.  }
   \label{fig:shift}
\end{figure*}

In real CPI-C observations, dedicated reference star observations may not be feasible due to time and resource constraints. Instead, flux calibration of the central star can be achieved by offsetting the target slightly from the coronagraphic mask and acquiring a set of short-exposure, non-coronagraphic images (as shown in Fig. \ref{fig:shift}). These direct observations of the stellar PSF can then be used to estimate the stellar flux and calibrate the planet-to-star contrast. This strategy allows for robust photometric referencing without requiring a separate reference star or coronagraph-off sequence.

\subsection{Photometric Analysis} \label{sec:phot}
Next, we apply aperture photometry to measure the magnitude of sources in the image. The \texttt{sep} code (\citealt{2016JOSS....1...58B}) is used to determine the exact coordinates of the target planet. Then, \texttt{photutils} (\citealt{larry_bradley_2024_10967176}) is employed to select a circular aperture with an appropriate radius for the target, and a square aperture for the background area, based on the position and size of the dark region. The combined flux within the target and background apertures, as well as the standard deviation within the background aperture, are calculated separately. The target aperture is used as a mask when measuring the background aperture to minimize contamination from the target.

The following parameters were obtained statistically: the standard deviation and the sum of readings in the background area, the signal at the target aperture, the number of pixels $a_0$ in the target aperture, and the number of pixels $a_s$ in the background area. Additionally, two known instrument parameters were used: the $\rm GAIN$, representing the number of photoelectrons per ADU (analog-to-digital unit), and the electron multiplication ($\rm EMGAIN$) occurring inside the EMCCD.

The background value is determined by averaging the readings in the skylight background area, i.e., $D_s/a_s$. Therefore, the number of photons in the target signal is the total signal within the aperture minus the background signal within the aperture:

\begin{equation}
S_0=K \cdot\left(D_0-D_s \cdot\left(a_0 / a_s\right)\right),   \label{eq:s0}
\end{equation}
where $K$ is the ratio of $\rm GAIN$ to $\rm EMGAIN$, it represents the number of photons corresponding to each ADU.

By using this method to measure the number of photons from both the planet and a reference star with the same aperture size, the magnitude of the planet can be calculated based on the known magnitude of the reference star. Subsequently, the observational contrast of the planet relative to the star can be determined.

To calculate the noise, we account for the noise generated by both the target source and the background, relative to the number of pixels in the chosen apertures:

\begin{equation}
\varepsilon\left(S_0\right)=\sqrt{S_0+\sigma_s^2 \cdot a_0 \cdot\left(1+a_0 / a_s\right)}.  \label{eq:error}
\end{equation}

The measurement error between the target planet and the reference star is calculated using Equation \ref{eq:error}, and the photometric error of the target planet is obtained through error propagation.

At this stage, the photometry of the image target is complete, and the contrast and signal-to-noise ratio of the planet have been determined. The photometry results for the four bands are presented in Table \ref{tab:phot_result}.

\begin{table}
\bc
\begin{minipage}[]{100mm}
\caption[]{The resulting contrast and signal-to-noise ratio of four bands. \label{tab:phot_result}}\end{minipage}
\setlength{\tabcolsep}{1pt}
\small
 \begin{tabular}{cccc}
  \hline\noalign{\smallskip}
Filter& Exposure& Contrast& SNR\\
 & (s) & ($\times 10^{-8}$) & \\
  \hline\noalign{\smallskip}
F520& $17\times 30$ & $5.163$ & 8.870 \\
F662& $9\times 30$ & $4.737$ & 4.451\\
F720& $50\times 30$ & $3.533$ & 5.480\\
F850& $100\times 30$ & $2.138$ & 5.488\\
  \noalign{\smallskip}\hline
\end{tabular}
\ec
\tablecomments{0.86\textwidth}{The photometric results of the target planet in the simulated image used as an example.}
\end{table}

\begin{figure} 
   \centering
   \includegraphics[width=12.0cm, angle=0]{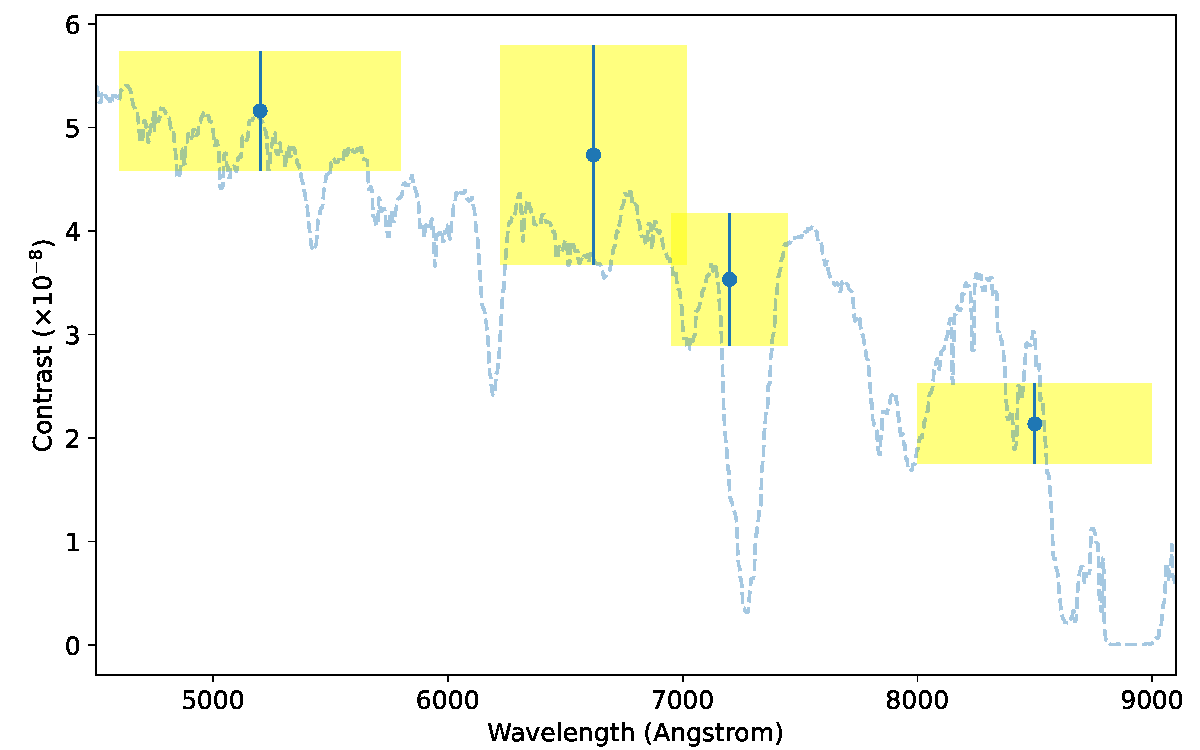}
   \caption{Comparison of photometric results and theoretical spectra.} 
   \label{fig:modelcompare}
\end{figure}

Fig. \ref{fig:modelcompare} shows the photometric results of the simulated planet in four bands alongside the theoretical planetary spectrum calculated based on the input parameters. The resemblence between the photometry of the simulated images and the theoretical models confirms that key physical parameters, such as planetary albedo and phase angle, were correctly implemented and processed.

\subsection{Exposure Time and EM Gain Effects}

\begin{figure} 
   \centering
   \includegraphics[width=12cm, angle=0]{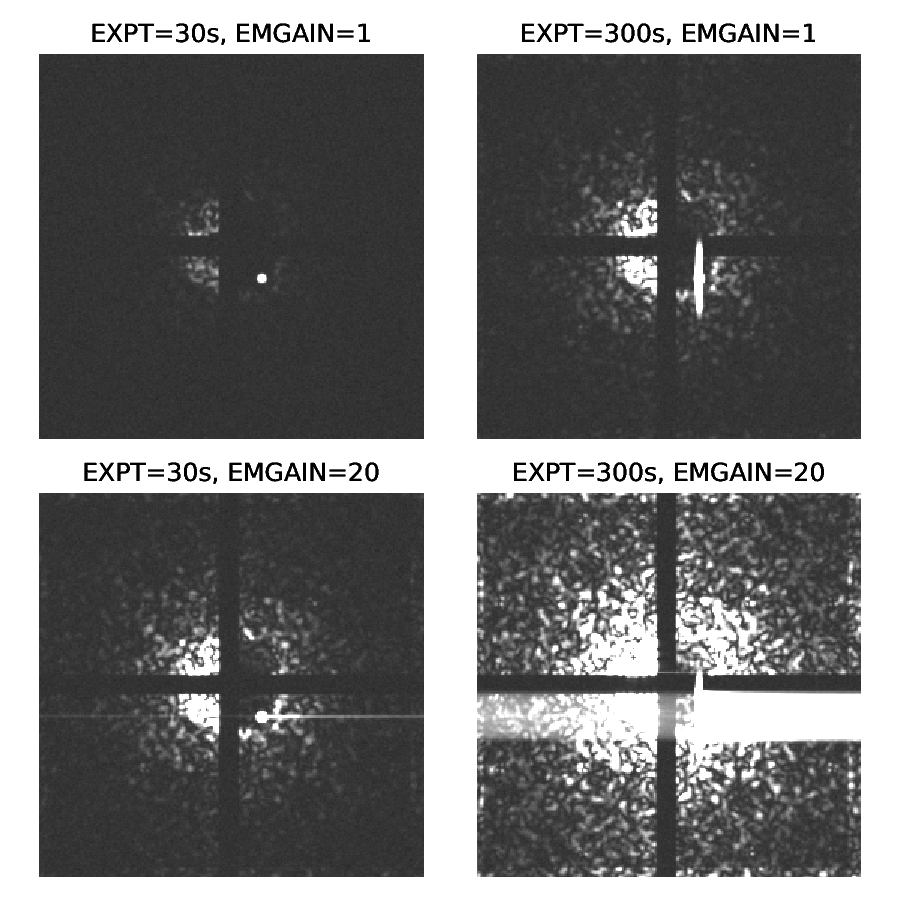}
   \caption{Simulated CPI-C images generated at different exposure times (30 s, 300 s) and EM gains (EMGAIN = 1, 20). A bright source is added in the dark hole on the lower right at a position angle of $135^\circ$ and an angular distance of $0.95^{\prime \prime}$. }
   \label{fig:effect}
\end{figure}

Exposure time and EM gain are critical observational parameters that significantly influence the performance of EMCCD-based coronagraphic imaging instruments. To evaluate and illustrate these effects within our \texttt{CPISM} simulations, we present simulated CPI-C images generated at different exposure times (30 s, 300 s) and EM gains (EMGAIN = 1, 20), as shown in Fig. \ref{fig:effect}.

The simulated results clearly demonstrate distinct trends. At a short exposure time (30 s) with low EM gain (EMGAIN = 1), the image is dominated by detector noise and shows limited sensitivity, with only the brightest targets barely detectable. Increasing the exposure time to 300 s at low EM gain significantly enhances signal detection but also leads to saturation and blooming in the brightest regions, resulting in characteristic vertical streak artifacts in the image.

When employing higher EM gain (EMGAIN = 20), even a relatively short exposure (30 s) greatly amplifies the photon-generated electrons, improving sensitivity to faint sources. However, the amplified noise and bright-source saturation artifacts become pronounced, demonstrating a trade-off between sensitivity and image fidelity. In the case of a longer exposure (300 s) combined with higher EM gain, severe saturation and blooming effects dominate the image, substantially degrading image quality and making accurate photometric extraction challenging or impossible.

\section{Summary}
\label{sec:dis}
In this paper, we introduce \texttt{CPISM}, the simulation program for CPI-C, one of the backend modules of the CSST. The program includes modules for generating stellar spectra, planetary albedo models, and contrast calculations. It enables users to input parameters for stars and planets, and to compute expected photometric and spectral outputs. Additionally, the program can calculate the required exposure time for observations or the signal-to-noise ratio of the observational results.

We also present a methodology for simulating CPI-C observation targets using a stellar spectrum model and a planetary albedo model to measure the contrast. Based on the modular framework, \texttt{CPISM} includes functional components for target modeling, imaging simulation, observational effects, and detector response. The complete simulation workflow is then described, from parameter configuration to Level 0 data generation, integrating optical propagation, PSF convolution, noise modeling, and EMCCD detector behaviors.

We use Alpha Centauri as an example to demonstrate the program’s capability to simulate realistic high-contrast imaging observations. By applying a RDI technique and aperture photometry, we extract contrast and signal-to-noise ratio values of a hypothetical planet across multiple bands. The photometric results show good agreement with the theoretical input spectra. The example demonstrate the end-to-end operability of the simulator—from parameter input to Level 0 data output—under a controlled, simplified observational scenario.

This result not only verifies the scientific robustness of our simulation but also provides a solid foundation for applying the program to future survey strategy. The close alignment between simulated and theoretical results indicates that the program can reliably predict observational outcomes under various conditions, highlighting its potential as a tool for optimizing high-contrast exoplanet observations. This accuracy is crucial for developing strategies that maximize observational efficiency and scientific return from instruments like CPI-C.

Further analysis of exposure time and EM gain demonstrates \texttt{CPISM}'s capability to accurately reproduce image artifacts and sensitivity variations. Shorter exposures and lower EM gains result in detector-noise dominated images, whereas longer exposures and higher gains increase sensitivity but also cause saturation and blooming artifacts. This analysis highlights \texttt{CPISM}'s practical utility in optimizing observational parameters.

Our simulation program is highly flexible. With the provided interfaces, the stellar spectrum model and planetary albedo model can be easily replaced. In this article, we have selected a relatively simplified atmospheric model as an example, but users can substitute it with a more sophisticated model based on their needs. Additionally, the observation band used in the simulation can be changed, provided a corresponding transmittance curve is available. This program will also assist in selecting optimal observation bands for CPI-C.

Looking ahead, this simulation framework will play a crucial role in advancing exoplanetary research. As the CPI-C module is poised for groundbreaking discoveries, the insights provided by our simulations will be instrumental in detecting and characterizing new worlds beyond our solar system.

\normalem
\begin{acknowledgements}
This work is based on the mock data created by the CSST Simulation Team, which is supported by the CSST scientific data processing and analysis system of the China Manned Space Project. We acknowledge National Natural Science Foundation of China (NSFC) under grant nos U2031210 and 11827804, as well as the science research grants CMS-CSST-2021-A11, CMS-CSST-2021-B04, CMS-CSST-2025-A17, CMS-CSST-2025-A18 and CMS-CSST-2025-A19 from the China Manned Space Project. This research is also funded by the ``Jiangsu Funding Program for Excellent Postdoctoral Talent''.

\end{acknowledgements}


\bibliographystyle{raa}
\bibliography{bibtex}

\end{document}